\documentclass[pra,superscriptaddress,groupedaddress,a4paper,twocolumn,nofootinbib]{revtex4}
\usepackage{amsmath,amsthm,amsfonts,amssymb,times,graphicx,color}
\usepackage{multirow}
\usepackage{subfigure}
\usepackage{mathtools}
\usepackage{url}
\usepackage{color}
\mathtoolsset{showonlyrefs}

\newcommand{\couic}[1]{}
\newcommand{\couicfootnote}[1]{}
\newcommand{\couicefootnote}[1]{}

\newcommand{\ket}[1]{| #1 \rangle}
\newcommand{\bra}[1]{\langle #1 |}
\newcommand{\Z}{\mathbb{Z}}

\def\OD#1{{\textcolor{black}{#1}}}

\begin{document}

\title{Coarse-grained quantum cellular automata}
\author{O. Duranthon}
\affiliation{Aix-Marseille Université, CNRS, Laboratoire d'Informatique et Systèmes, Marseille, France and Département de Physique, École Normale Supérieure, Paris, France}
\author{Giuseppe Di Molfetta}
\email{giuseppe.dimolfetta@lis-lab.fr}
\affiliation{Aix-Marseille Université, CNRS, Laboratoire d'Informatique et Systèmes, Marseille, France}

\date{\today}

\begin{abstract}
One can think of some physical evolutions as being the emergent-effective result of a microscopic discrete model. Inspired by classical coarse-graining procedures, we provide a simple procedure to coarse-grain color-blind quantum cellular automata that follow Goldilocks rules. The procedure consists in (i) space-time grouping the quantum cellular automaton (QCA) in cells of size $N$; (ii) projecting the states of a cell onto its borders, connecting them with the fine dynamics; (iii) describing the overall dynamics by the border states, that we call signals; and (iv) constructing the coarse-grained dynamics for different sizes $N$ of the cells. A byproduct of this simple toy-model is a general discrete analog of the Stokes law. Moreover we prove that in the spacetime limit, the automaton converges to a Dirac free Hamiltonian. The QCA we introduce here can be implemented by present-day quantum platforms, such as Rydberg arrays, trapped ions, and superconducting qbits. We hope our study can pave the way to a richer understanding of those systems with limited resolution. 
\end{abstract}

\maketitle

\section{Introduction}

Cellular automata (CA) are discrete dynamic systems whose rules appear very simple, but whose emerging phenomenology is complex \cite{wolfram1984cellular}. An example is Conway's famous game of life \cite{adamatzky2010game}: simple understandable rules produce an entire animated world, with blinkers, gliders, guns and Garden of Eden states. Historically, they have been adapted to describe several complex systems such as hydrodynamic fluids \cite{rothman2004lattice}, traffic patterns \cite{kerner2002cellular}, the formation of biological processes \cite{ermentrout1993cellular} and reaction-diffusion \cite{modSysPh}. One of the first definitions of cellular automata is that of von Neumann \cite{von2012computer}, whose rules are applied locally on a two-dimensional grid, in discrete time. It soon became clear that many CA are universal models of computation, or in other words they effectively simulate Turing machines \cite{turingCompl}. More formally CA are a discrete set of cells whose values are in identical finite sets; a state is the current value of the cells; cells evolve in discrete time according to a local translation-invariant update rule.

Although CA describe a very broad spectrum of classical phenomena, they cannot describe a quantum system. In order to do so correctly and effectively, to the best of our knowledge today, we must consider the quantum analog of CA, namely quantum cellular automata (QCA). QCA can be defined in various ways, and many of them are recently proved equivalent \cite{revueA}. We define our QCA over a general graph where in each vertex sits a cell, with each cell composed of sub-cells defined as finite-dimensional quantum systems. The evolution of the QCA is given by local unitaries that act on partitions of the graph. From the above definition it is clear that all QCA in this model are reversible, as only unitary operations are performed. This is in accordance with the expected microscopic, fully quantum dynamics.
These automata can be used as a model of distributed quantum computation and in particular they can provide quantum schemes to simulate quantum physics theories \cite{arrighi2018dirac, di2013quantum,di2016quantum, arnault2016quantum}. Indeed QCA offer a natural framework to theoretical physics. They seem promising since they satisfy fundamental concepts such as unitarity and causality, and we can make them gauge invariant \cite{invJaugeCA, qedArrighi, arnault2019discrete} and even Lorentz covariant \cite{invLorentz, bisio2017quantum, debbasch2019discrete}.

\OD{If a quantum description of a discrete dynamic system does not pose major problems, the transition to classical is not yet free of difficulties. In order to study such transition in the discrete framework of the QCA, here we imagine that a hypothetical ”detector” is not good enough to resolve the more basic level of the dynamic, which translates into a loss of information. This can be formalized by a quantum coarse-graining procedure, namely a completely positive trace
preserving (CPTP) map. The idea to coarse-grain complex systems other than particles or polymers is very recent, and range from quantum chemistry \cite{grChQu} and high energy physics \cite{bartlett07,oleg,grQF} to biology \cite{fontana}. Few studies have been produced in the context of quantum walks \cite{d2016virtually, bisio2016quantum} and optical lattices \cite{correia2019spin}
In this article, inspired by classical coarse graining procedures \cite{israeli2006coarse, costa2019coarse} and a first recent result for quantum systems \cite{duarte2017emerging}, we space-time group unit cells in one supercell, and we construct effective emergent dynamics for different sizes of the supercell. In the  lowest level we have the fully microscopic and reversible quantum dynamics; while for increasing supercell size the quantum features are gradually suppressed and a classical \OD{dynamics emerges}}.

\OD{The use of QCA is particularly useful for a coarse-graining procedure, as its discrete structure gives a clear visualization of the process: instead of being able to resolve a single cell, one is only able to see bigger blocks, thus not taking into account the full information about the microscopic structure. This erasure of information leads to an effective state and effective discrete dynamics, which is possibly simpler than the microscopic fully quantum dynamics and thus, at some level, likely to be simulated in an efficient way by a classical computer.}

\OD{In order to explore this idea, we specialize to a very broad class of QCA, following the so-called Goldilocks rules, recently introduced in \cite{montangero}. The Goldilocks rules are trade-offs of the kind underpinning biological, social, and economic emergent complexity. For the sake of simplicity, we investigate the simplest, but still non trivial, rule which makes the state evolution depending on a neighborhood of unitary radius.}

\OD{The interest for such a choice is double: (i) as is proven in Appendix \ref{colourBlindness}, this Goldilocks QCA is color-blind, i.e. is global invariant under the change $0\leftrightarrow 1$; (ii) the overall dynamics may be described by border states, we call \textit{signals}, leaving on the edges of the grid. We will see that these signals are natural candidates for describing the QCA at different level description}.\\

\OD{The article is organized as follows. In section \ref{GQCA} we define the Goldilocks QCA. Then in section \ref{decimationAndCG} we introduce the coarse-graining procedure and we derive the dynamical law for a given size of the supercell. We conclude and discuss in section \ref{sec:conclusion}. Moreover we provide two appendices: in Appendix \ref{colourBlindness} we give a full rigorous proof of the color blindness of the Goldilocks QCA; in Appendix \ref{stokes} we show an interesting by-product of the coarse-graining procedure, namely, a discrete analogous of the Stokes law.}

\section{Goldilocks rules}
\label{GQCA}

\begin{figure}[h!]
 \centering
 \includegraphics[scale=0.4]{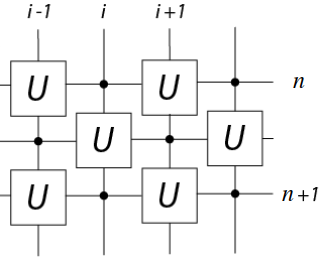}
 \caption{The GQCA in one spatial dimension. Squares represents local unitary gates $U$ applied to a qbit state.}
 \label{portesU}
\end{figure}

The Goldilocks QCA  (GQCA) is defined over a one-dimensional qbit string as depicted in Fig. \ref{portesU}. Each qbit $\ket{e_i} \in \mathbb{C}^2$, updates in discrete time steps, according to a local gate defined on the qbit’s neighborhood as follows:
\begin{equation}
U_i = \sum_{\sigma=e_{i-1},e_{i+1}} |\sigma\rangle\langle \sigma|\otimes \OD{V_i^{c_\sigma}}.
\end{equation}
acting on the qbit at position $i$, where the sum is performed on the four possible configurations $\sigma$ of the two neighbors of $i$. The parameter $c_\sigma\in \{1,0\}^2$ controls whether $V_i$ is applied or not to $|e_i\rangle$. According to the simplest Goldilocks rule $T_6$ \cite{montangero}, we have to apply $V$ to $|e_i\rangle$ if its two neighbors are different, otherwise we have to keep it unchanged, leading to the following rule: $c_{00}=0$, $c_{11}=0$, $c_{01}=1$ and $c_{10}=1$.
\begin{figure}[h!]
 \centering
 \includegraphics[scale=0.25]{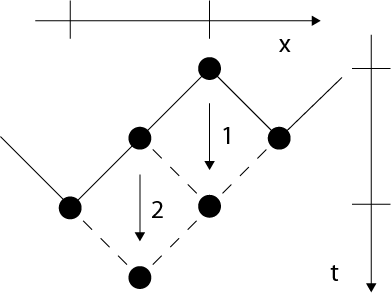}
 \includegraphics[scale=0.25]{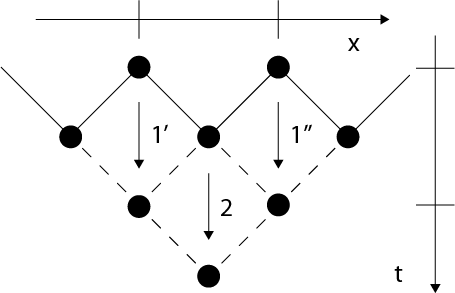}
 \includegraphics[scale=0.35]{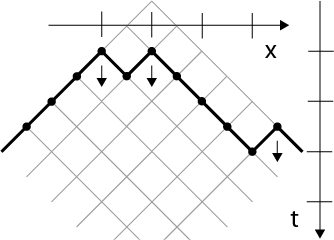}
 \caption{\label{evLigne} The looser evolution rule. \emph{\OD{Top left}} we can evolve 1, via unitary transformation $U_1$, then 2, via $U_2$. \emph{\OD{Top right}} we can evolve $1'$, via $U_{1'}$ and $1"$, via $U_{1"}$, then 2, via $U_{2}$. \emph{Bottom} the bold line is the current state.}
\end{figure}
The dynamics of the GQCA strongly depends on the choice of $V$. A sufficiently general expression for $V$ is the following: 
\begin{equation}
\label{eq:V}
V = e^{i\phi}[\cos(m\varepsilon)\sigma_x+\sin(m\varepsilon)\sigma_z],
\end{equation}
where $\phi \in [0,2 \pi[$ and $m$ is a constant real parameter, which we call the $\textit{mass}$. The real parameter $\varepsilon$ is the characteristic length of the space-time grid. 

\begin{figure}[h!]
 \centering
 \includegraphics[scale=0.5]{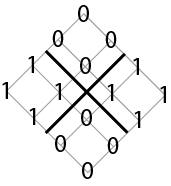}
 \caption{\label{domaines} The time evolution of the GQCA and the domain walls (bold lines) with $\phi=0$ and $m=0$.}
\end{figure}

\begin{figure}[h!]
 \centering
 \includegraphics[scale=0.22]{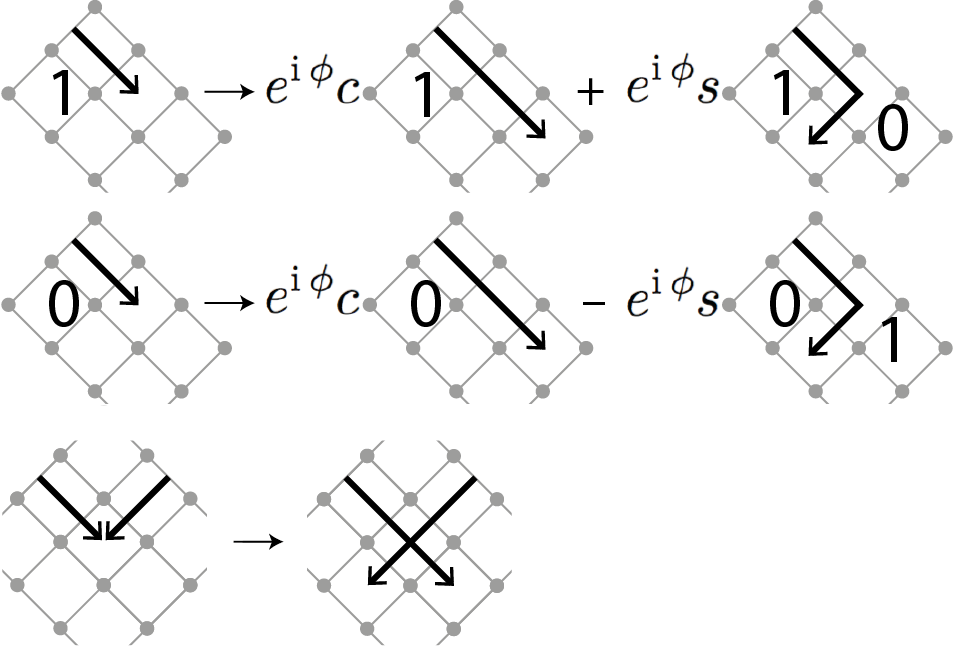}
 \caption{\label{transitions} One time step evolution rules of $\bar\psi$. \OD{The two upper lines are for one $\bar\psi^+$; $\bar\psi^-$ is mirrored.}}
\end{figure}

One may remark that the scheme depicted in Fig.\ref{portesU}, still well defined, does not put time and space on an equal footing, because we apply first gates at odd positions at step $n$ and later gates at even positions at step $n+\frac{1}{2}$. To avoid this inconvenience, we can equivalently represent the GQCA, depicted in  Fig.\ref{portesU}, as in Fig.\ref{evLigne}. Let us introduce the position $x_i = \frac{i}{2} \varepsilon$ and the time $t_n =n \varepsilon$. Since the unitaries $U_i$ and $U_j$ commute if $|i-j|>1$, we say that the state at position $x_i$ and time $t_n$ can evolve to the position $x_i$ at time $t_{n+1}$, if its two neighbors at positions $x_{i-1}$ and $x_{i+1}$ are ahead at time $t_{n+1/2}$. In Fig.\ref{evLigne}, the black solid arrows represent the local time evolution of the state and the current state is then represented by a broken solid line.
The dynamics of a GQCA locally looks very simple, as is shown in Figs.\ref{domaines} and \ref{transitions}, although at large scale it may lead to high complexity \cite{montangero}. In fact, by looking at its spacetime diagram, we can observe finite regions of $\ket{0}$'s and $\ket{1}$'s propagating quite regularly. In this scenario, the \textit{domain walls} between these regions, appear to be the most relevant point to investigate. This suggests a dual representation of the GQCA.
Let us introduce an auxiliary state $\psi \in \mathbb{C}^2$, sitting on the edges of the space-time grid: this new discrete field is defined in order to account, modulo $2$, the local difference of the two nearest neighbor qbits. If they are different, $\psi=\ket{1}$, otherwise $\psi=\ket{0}$. More formally, taking $(x_i, t_n)$ and $(x_j, t_m)$ neighbors, $(x, t)$ being the midpoint:
\begin{equation}
\psi(x, t) = \ket{e(x_i,t_n)\oplus e(x_j,t_m)}.
\end{equation}

\begin{figure}[h!]
 \centering
  \includegraphics[scale=0.4]{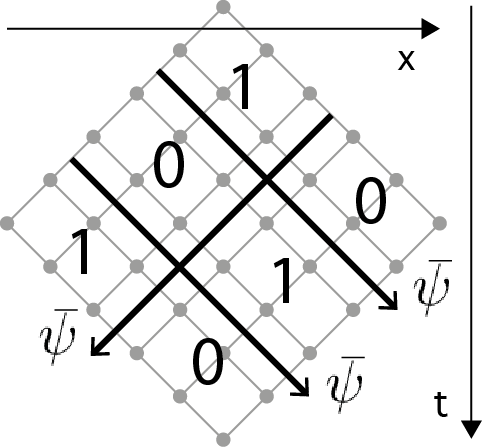}
  \caption{\label{evPsis} The time evolution of the $\bar \psi$'s (bold lines)  with $\phi=0$ and $m=0$.}
 \end{figure}

As we can see in Fig. \ref{evPsis}, the $\bar\psi\equiv\ket{1}$ \OD{propagates} in space-time at speed $1$, following the domain walls that separate regions of $\ket{0}$'s and $\ket{1}$'s. 

Indeed, we can fully describe the automaton only looking at the signals $\bar\psi$. Each of them behaves like a quantum walker over the dual grid. If we look at the one $\bar \psi$-subspace, a straightforward calculation leads us to the following recursive relations:
\begin{small}
\begin{equation}
\begin{split}
\bar\psi^+(x,t+\frac{\varepsilon}{2})=\OD{e^{i\phi}}\left[c_\varepsilon\bar\psi^+(x-\frac{\varepsilon}{2},t)\pm s_\varepsilon\bar\psi^-(x,t)\right]\\ \bar\psi^-(x,t+\frac{\varepsilon}{2})=\OD{e^{i\phi}}\left[c_\varepsilon \bar\psi^-(x+\frac{\varepsilon}{2},t)\mp s_\varepsilon \bar\psi^+(x,t)\right]
\end{split}
\label{eq:receq}
\end{equation}
\end{small}
\OD{where we set $c_\varepsilon = \cos(m\epsilon)$ and $s_\varepsilon = \sin(m\epsilon)$.}
The $\bar\psi^+$ (respectively $\bar\psi^-$) is the right- (respectively left-)moving signal and the sign choice depends on whether we have $0-\bar\psi-1$ or $1-\bar\psi-0$.

\OD{However, as we rigorously prove in Appendix \ref{colourBlindness}, the automaton is color-blind: \emph{the global time evolution of the $\bar\psi$ is invariant by the flip $\ket{0} \leftrightarrow \ket{1}$.}}

The constant $\phi$ selects the signal statistics: when two $\bar\psi$'s cross, we obtain a phase delay of $2\phi$ and this phase induces a statistics. For $\phi=\pi/2$, the $\bar\psi$'s have to be considered fermions. The local evolution rules for $\bar\psi$ are summarized in Fig. \ref{transitions}.

\OD{The fact that the signals $\bar\psi$ actually behave as (Dirac) quantum walkers in ($1+1$) is not surprising: quantum walkers are usually though as the one-particle sector of a QCA. In this context, the domain walls, although defined in a multiparticle setting, behave geometrically as one-dimensional states. Moreover, looking at the single-particle sector and gauging the global phase away, we can take the continuous limit of \eqref{eq:receq}. Expanding around $\varepsilon=0$ of Eq. \eqref{eq:receq} and taking formally the limit for $\varepsilon\to 0$, the single particle dynamics recovers the Dirac-like equation in ($1+1$) space-time dimensions:
\begin{equation}
   i \partial_t \tilde\psi = H_D \tilde\psi,
    \label{eq:DE}
\end{equation}
where $\tilde\psi = (\psi^+,\psi^-)$ and $H_D = i \sigma_z \mp 2m\sigma_y$.}


\section{Decimation and coarse-graining}
\label{decimationAndCG}

\OD{The procedure to coarse-grain the above automaton is decomposed in three steps. First we partition the space-time diagram in cells of size $N$. Then we sample a number of qbits on the borders of these cells, and we refer to them as probes. Finally we coarse-grain by a CPTP map the signals, constructing the effective dynamics for different sizes of the cells. Notice that as a by-product of the coarse-graining procedure, we can always establish a connection between the probes, along the domain walls and the signals, traveling through the cells, as we prove in Appendix \ref{stokes}. }

\begin{figure}[h!]
 \centering
 \includegraphics[scale=0.3]{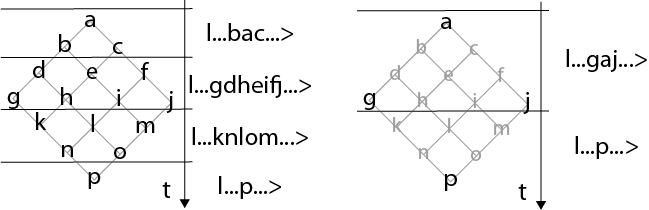}
 \includegraphics[scale=0.3]{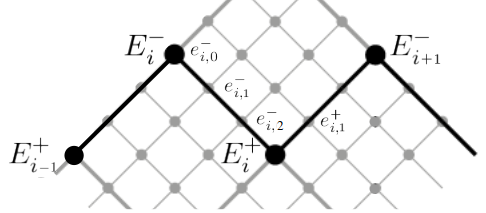}
 \caption{\label{grgr} \emph{Top:} evolution and states without and with graining. The initial states are the upper ones (g, d, b, a, c, f and j). $N=3$. \emph{Bottom:} the bold line is the current coarse-grained state. }
\end{figure}

\OD{Let us first illustrate this procedure in general terms. Consider a space-time diamond of size $N$ and a line of initial qbits, arranged as in Fig. \ref{grgr}, its extremities being fixed. From a microscopic point of view, the unitary evolution of one qbit $\ket{e_{i}}$ is given locally by $U_{i}$. Now let us imagine that an available detector is not able to resolve the microscopic dynamics within a space-time super-cell of size $N$, but may only have access to a high-level description. Here, we assume that such macroscopic picture is the result of a space-time sampling of the automaton onto some sites. Then, we retro-engineer a CPTP map in order to recover such probes from the bottom. Without lack of generality, we can assume that such probes are sampled onto the vertices of a space-time diamond of size $N$, as depicted in Fig.\ref{grgr} (top-right).}
\OD{If we project onto the probes by brute force, at each stroboscopic time $t=nN$, we would have access to a state of the form:
\begin{equation}
\tilde \rho = \bigotimes_{i \in \Z} |E^-_i E^+_i\rangle\langle E^-_i E^+_i|
\end{equation}
where $E^-_i = \ket{e^-_{i, 0}}$ and $E^+_i=\ket{e^+_{i, 0}}$. 
We mean by brute force, killing all the fine qbits between the coarse qbits, i.e. all $\ket{e^-_{i,j}}$ and $\ket{e^+_{i,j}}$ for $j=1\ldots N-1$. We ought to do better than that, trying to preserve the maximum of coherence shared among the fine states. }

\OD{In order to formally build such a quantum channel, in the following we consider the one-particle sector of the dual GQCA, illustrated in Sec. \ref{GQCA}. }


\subsection{Non-interacting signals}

Let us start to consider a fine state composed by two qbits $|ef\rangle$ and its density matrix $\rho =|ef\rangle\langle ef|$. If our detector has a finite resolution, we may expect that two close enough qbits are not well discriminated; for instance, it could not be good enough to distinguish the states $|01\rangle$ and $|10\rangle$. At a higher level both states could be mapped to the state $|1\rangle$. Such a coarse-graining is somehow arbitrary but in the case in which the channel is quantum, the map has to be trace preserving and completely positive \cite{nielsen2002quantum}. The first condition translates in demanding that the population terms are preserved: 
\begin{eqnarray*}
	\Lambda\left(|00\rangle\langle00|\right)&=&|0\rangle\langle0|~;\\
	\Lambda\left(|01\rangle\langle01|\right)&=&|1\rangle\langle1|~;\\
	\Lambda\left(|10\rangle\langle10|\right)&=&|1\rangle\langle1|~;\\
	\Lambda\left(|11\rangle\langle11|\right)&=&|1\rangle\langle1|~.
		\label{eq:rules_po}
\end{eqnarray*}
Let us consider that coherence terms are transformed as follows:
\begin{eqnarray*}
	\Lambda\left(|01\rangle\langle00|\right)&=&a|1\rangle\langle0|~;\\
	\Lambda\left(|10\rangle\langle00|\right)&=&a|1\rangle\langle0|~;\\
	\Lambda\left(|11\rangle\langle00|\right)&=&a|1\rangle\langle0|~,
	\label{eq:rules_co}
\end{eqnarray*}
with $a$ being a constant that we aim to constrain. The associated Choi matrix is a block matrix, with blocks $(i,j)$, $\Lambda(|i\rangle\langle j|)$, $i,j$ being the element of the computational basis: 
\begin{equation}
\begin{split}
C_\Lambda&=\begin{pmatrix}
\begin{pmatrix}1&0\\0&0\end{pmatrix}&\begin{pmatrix}0&a\\0&0\end{pmatrix}&\begin{pmatrix}0&a\\0&0\end{pmatrix}&\begin{pmatrix}0&a\\0&0\end{pmatrix} \\
\begin{pmatrix}0&0\\a^*&0\end{pmatrix}&\begin{pmatrix}0&0\\0&1\end{pmatrix}&0&0 \\
\begin{pmatrix}0&0\\a^*&0\end{pmatrix}&0&\begin{pmatrix}0&0\\0&1\end{pmatrix}&0 \\
\begin{pmatrix}0&0\\a^*&0\end{pmatrix}&0&0&\begin{pmatrix}0&0\\0&1\end{pmatrix}
\end{pmatrix}\\
\end{split}
.
\label{eq:Choi}
\end{equation}
Its non-null eigenvalues are $\lambda=1,1,1\pm\sqrt{3aa^*}$~ and they must be positive. Consequently, 
\begin{equation}
    |a|\le 1/\sqrt 3,
\end{equation}
which defines an upper bound for the coherence coefficients. Note that the choice $a=0$ translates in a projection onto the coarse states.

If we consider a more general coarse-graining onto $L$ coarse states, where each coarse state corresponds to $N$ fine states, matrix \eqref{eq:Choi} reads
\begin{equation}
C_\Lambda\simeq\begin{pmatrix}
\mathbf 1_N&\mathbf a&\cdots&\mathbf a\\
\mathbf a^*&\mathbf 1_N&\cdots&\mathbf a\\
\vdots& &\ddots & \\
\mathbf a^*&\mathbf a^*&\ldots&\mathbf 1_N
\end{pmatrix}.
\end{equation}
The Choi matrix is now a block matrix whose $L$ blocks have dimension $N\times N$ and $\mathbf a$ is a block filled with $a$. Notice that we choose this particular form because we would like a $\Lambda$ which is invariant by permutation or translation of the coarse-states.

Again it must be positive. In particular, for $a=0$ its eigenvalues are all positive; to study the general case, we use their analyticity in $a$ and we look for a null eigenvalue; this is equivalent to studying the $L\times L$ matrix: 
\begin{equation}
\begin{pmatrix}
1&Na&\cdots&Na\\
Na^*&1&\cdots&Na\\
\vdots& &\ddots & \\
Na^*&Na^*&\cdots&1
\end{pmatrix}.
\end{equation}
After simple algebra, it is straightforward to verify that the spectrum of the above matrix is positive for a small set of $a(N)$ encompassing 0, and that the maximum value coincides with $a_\mathrm{max}=1/N.$

Let $$\rho = \sum_{i,j} c_{ij}|s_i\rangle\langle s_j|$$ be a fine grained density matrix and $\Lambda$ be the uniform coarse-graining that gives the largest coherence; then
\begin{small}
\begin{equation}
\rho\xrightarrow{\text{$\Lambda$}} \sum_{i,j} \tilde c_{ij}|\Lambda s_i\rangle\langle\Lambda s_j|,\quad \tilde c_{ij}=
\begin{cases}c_{ij} & \mathrm{if\,}\Lambda s_i=\Lambda s_j,\\ c_{ij}/N &\mathrm{otherwise.}\end{cases}
\label{eq:CGmap1}
\end{equation}
\end{small}
This shows that coarse-graining a quantum state reduces coherence and, then, the quantumness of the system. One can estimate this loss: coherence is divided by the cardinal $N=\#\Lambda^{-1}$; it is as if the coherence of the coarse state were shared among its $N$ fine states. Notice that this does not depend on $L$ and this is also composable with respect to the iteration of coarse-graining. 

\begin{figure}[h!]
 \centering
  \includegraphics[scale=0.3]{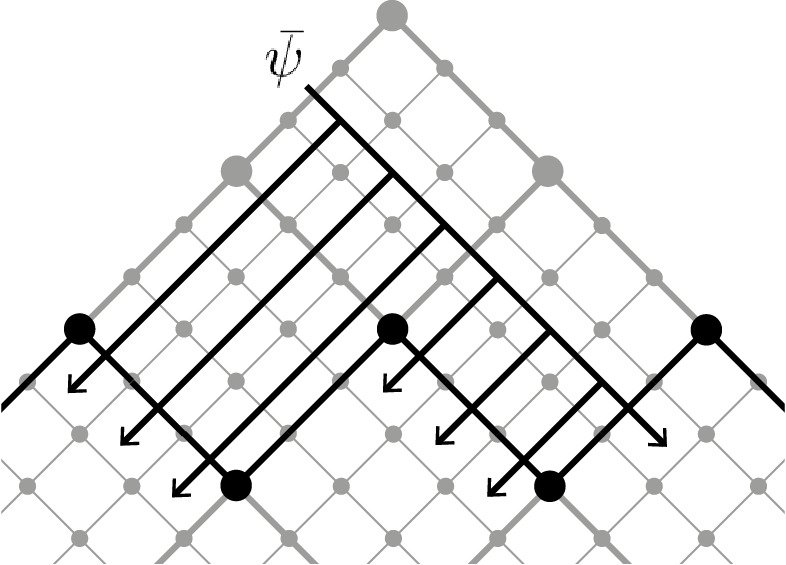}
  \includegraphics[scale=0.3]{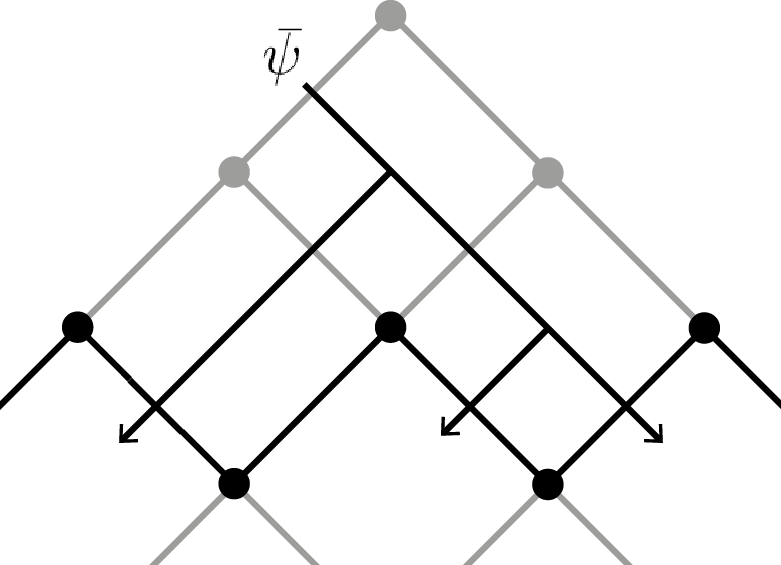}
  \caption{\label{grgrPsiDirac} Coarse-graining of a single $\bar\psi$ at order $\epsilon$. $N=3$, $n=1$. The many arrows should be understood as a superposition of states. $N$ left moving states are mapped onto one coarse state.}
\end{figure}

Finally, we may also interpret the above result as the $N$ possible positions a signal $\bar\psi$ can take going through two coarse qbits: we cannot distinguish these trajectories as depicted in Fig. \ref{grgrPsiDirac}.


\OD{Notice that, in the formal limit for $N\rightarrow\infty$, the dynamics describes a fully classical cellular automata, where the right- and left-moving coarse signals are completely decoupled.}

\OD{The most interesting case is for finite $N$, where the effect of this coarse-graining map translates in renormalizing the coupling parameter $m$, embedded in the coherence terms of the density matrix, by a factor $N$. In the following we focus on the case for $\varepsilon \ll 1$. Let us consider the evolved state in Eq.\eqref{eq:receq}, with $\phi=0$, starting from a single $\bar\psi^+$ localized at fine position $j$. At first order in $\varepsilon$, at time $nN$, the evolved state is:
\begin{equation}
\ket{\bar\psi}(nN) = \ket{\bar\psi^+_{n,j}}+\varepsilon m \sum_{i'=1-n}^{n}\sum_{j'=0}^{N-1} \ket{\bar\psi^-_{i',j'}}
\label{eq:evolvedPsi}
\end{equation}
where in the $0-1$ gauge
\begin{small}
\begin{equation}
\begin{split}
\ket{\bar\psi^+_{i,j}}=\bigotimes_{i'<i, j'}\ket{e^-_{i',j'}=0, e^+_{i',j'}=0} \bigotimes_{i'>i, j'}\ket{e^-_{i',j'}=1, e^+_{i',j'}=1}\\
\bigotimes_{j'}\ket{e^-_{i,j'}=0} \otimes \ket{e^+_{i,0}\ldots e^+_{i,j}e^+_{i,j+1}\ldots e^+_{i,N-1}=0\ldots 01\ldots 1}
\end{split}
\end{equation}
\end{small}
\begin{small}
\begin{equation}
\begin{split}
\ket{\bar\psi^-_{i,j}}=\bigotimes_{i'<i, j'}\ket{e^-_{i',j'}=0, e^+_{i',j'}=0} \bigotimes_{i'>i, j'}\ket{e^-_{i',j'}=1, e^+_{i',j'}=1}\\
\otimes \ket{e^-_{i,0}\ldots e^-_{i,j}e^-_{i,j+1}\ldots e^-_{i,N-1}=0\ldots 01\ldots 1} \bigotimes_{j'} \ket{e^+_{i,j'}=1}.
\end{split}
\end{equation}
\end{small}}
\OD{Now, the coarse-grained evolved density matrix reads:
\begin{small}
\begin{eqnarray}
\label{eq:evolvedDM}
\lefteqn{\Lambda \ket{\bar\psi}\bra{\bar\psi}(nN)} \\
 &=&\Lambda \ket{\bar\psi^+_{n,j}}\bra{\bar\psi^+_{n,j}}\\
 & &+\varepsilon m\sum_{i'=1-n}^{n}\sum_{j'=0}^{N-1} \Lambda(\ket{\bar\psi^-_{i',j'}}\bra{\bar\psi^+_{n,j}}+\ket{\bar\psi^+_{n,j}}\bra{\bar\psi^-_{i',j'}}) \\
 &=& \ket{\bar\Psi^+_n}\bra{\bar\Psi^+_n} + \varepsilon m\sum_{i'=1-n}^{n}N\frac{1}{N} (\ket{\bar\Psi^-_{i'}}\bra{\bar\Psi^+_n}+\ket{\bar\Psi^+_n}\bra{\bar\Psi^-_{i'}}) \\
 &=& \ket{\bar\Psi}\bra{\bar\Psi}(n)
\end{eqnarray}
\end{small}
with
$$
\ket{\bar\Psi}(n) = \ket{\bar\Psi^+_n}+\tilde\varepsilon m^\mathrm{cg} \sum_{i'=1-n}^{n} \ket{\bar\Psi^-_{i'}}\;,
$$
$$
m^\mathrm{cg} = \frac{ m}{N}\;,\quad\quad\tilde\varepsilon=N\varepsilon
$$
and
\begin{small}
\begin{equation}
\ket{\bar\Psi^+_{i}}=\bigotimes_{i'\leq i}\ket{E^-_{i'}=0, E^+_{i'}=0} \bigotimes_{i'>i}\ket{E^-_{i'}=1, E^+_{i'}=1},
\end{equation}
\end{small}
\begin{small}
\begin{equation}
\begin{split}
\ket{\bar\Psi^-_{i}}=\bigotimes_{i'<i}\ket{E^-_{i'}=0, E^+_{i'}=0}\bigotimes_{i'>i}\ket{E^-_{i'}=1, E^+_{i'}=1}\\
\otimes \ket{E^-_{i}=0, E^+_{i}=1}.
\end{split}
\end{equation}
\end{small}
The new space-time length is $\tilde\varepsilon$, which is consistent with the fact that the probes, located on the vertices of the space-time super-cell are far $N\varepsilon$. Moreover, from the above equation, it is clear to recognize the very same action of the local gate $V$, but with a renormalized mass $m^\mathrm{cg}$, on the coarse-grained state space. The dynamics is simple; the coherence is caught by the mass. As one can notice, at first order in $\varepsilon$, this coarse-graining map leaves essentially intact the dynamics up to a renormalization factor. On the new lattice with characteristic length $\tilde \varepsilon$, the formal continuous limit leads to a Dirac-like equation, as in \eqref{eq:DE}, with a renormalized mass $m^\mathrm{cg}$.  }

\section{Conclusion}
\label{sec:conclusion}

We studied a color-blind QCA, following the Goldilocks rules and we characterized its dual introducing extra qbits on the edges of the spacetime diagram. Such states play the role of discrete gradient, accounting the difference, modulo $2$, between neighbors qbits. We called them signals. Such entities propagate in space-time and coincide formally with quantum walkers, living on the borders between region of different color. A procedure to coarse-grain the overall microscopic system has been introduced. We first partition the space-time diagram in cells of size $N$ and we project each cell on some probes on the borders. The choice of the probes induces a unique quantum channel which coarse-grains the dynamics. We showed that the emergent automaton essentially behaves as the microscopic one with a renormalised mass, which scales inversely with the size of the cells, or in other terms with the resolution of the detector. The coherence shared among the microstates vanishes asymptotically with the size $N$ of the supercells.

Moreover, we found that it is possible to formally connect the coarse states (the probes) to the signal dynamics, connecting them by a discrete analogous of the Stokes law. This yields a rigorous geometrical correspondence between the (edge) coarse-states and the microscopic (bulk) dynamics, which, we believe, deserves further investigations.

\OD{From a purely theoretical point of view, the direction is to generalize the previous results to QCA with finite interactions and large mass ($m \epsilon\approx 1$), as in \cite{qedArrighi}. We suspect that, once the interaction is turned on, the emergent dynamics will lead to non-linearities \cite{correia2020macro} or diffusion terms \cite{costa2020quantum}. Moreover, the purpose is to provide new quantum simulation schemes for quantum field theories, taking into account the limited access that the observer has to the microscopic substrate. Also we may wonder whether the above results could be extended to higher dimensional space and in case of gauge field interaction. We leave it to future investigations.}  \\

\section{Acknowledgements} The authors acknowledge inspiring conversations with Fernando de Melo, Pedro Costa who contributed to spark the idea of the coarse-graining map in the context of Dirac QCA, and Pablo Arrighi and Nathanael Eon for inspiring discussions on symmetries and invariances in QCA. This work has been funded by the Pépinière d’Excellence 2018, AMIDEX fondation, project DiTiQuS and the ID 60609 grant from the John Templeton Foundation, as part of the "Quantum Information Structure of Spacetime (QISS)” project.

\appendix

\section{Color blindness}
\label{colourBlindness}

Color-blind cellular automata have been extensively studied in \cite{salo2013color}, where all cells get transformed by the same group element. The quantum analog has not been studied to the best of our knowledge. Although a complete characterization of this class of automata goes beyond the scope of this article, we will demonstrate in this section that our GQCA is color-blind and therefore subject to a global gauge invariance under transformation $0\leftrightarrow 1$.

Let us consider the gates
$$U_i = \sum_{s=e_{i-1},e_{i+1}} |s\rangle\langle s|\otimes V_i^{c_s},$$
with the same Goldilocks rules for $c_s$ we introduced in the main part of the article. The most general unitary for the gate $V$ reads
$$V = \begin{pmatrix}
a& b\\
-e^{i\phi}b^*& e^{i\phi}a^*
\end{pmatrix}$$
with $a=|a|e^{i\alpha}$, $b=|b|e^{i\beta}$, $|a|^2+|b|^2=1$ and $\phi$ real.

We compare the evolution of the particles $\bar \psi$ in the two possible gauges. Either the gauge is $0-\bar \psi-1$ and the evolution is
$$\begin{smallmatrix}
0& &1& &1\\
 &0& &1& \end{smallmatrix} \xrightarrow{\text{$U$}}
b\, \begin{smallmatrix}
0& &1& &1\\
 &0& &1& \\
0& &0& &1\end{smallmatrix}
+e^{i\phi}a^*\, \begin{smallmatrix}
0& &1& &1\\
 &0& &1& \\
0& &1& &1\end{smallmatrix}~;$$
or it is $1-\bar \psi-0$ and the evolution is
$$\begin{smallmatrix}
1& &0& &0\\
 &1& &0& \end{smallmatrix} \xrightarrow{\text{$U$}}
-e^{i\phi}b^*\, \begin{smallmatrix}
1& &0& &0\\
 &1& &0& \\
1& &1& &0\end{smallmatrix}
+a\, \begin{smallmatrix}
1& &0& &0\\
 &1& &0& \\
1& &0& &0\end{smallmatrix}.$$

Setting $\varepsilon=2$ for simplification, we summarize these rules as:
$$0-\bar \psi^+(x,t)-1 \rightarrow b\bar \psi^+(x+1,t+1)+e^{i\phi}a^*\bar \psi^-(x-1,t+1),$$
$$0-\bar \psi^-(x,t)-1 \rightarrow a\bar \psi^+(x+1,t+1)-e^{i\phi}b^*\bar \psi^-(x-1,t+1)~;$$
or in the second gauge
$$1-\bar \psi^+(x,t)-0 \rightarrow -e^{i\phi}b^*\bar \psi^+(x+1,t+1)+a\bar \psi^-(x-1,t+1),$$
$$1-\bar \psi^-(x,t)-0 \rightarrow e^{i\phi}a^*\bar \psi^+(x+1,t+1)+b\bar \psi^-(x-1,t+1).$$

We see that the change $0\leftrightarrow 1$ induces only phase shifts on transition rates and consequently preserves their amplitudes. There are only four possible phase shifts if we wish to change the gauge from $0-\bar \psi-1$ to $1-\bar \psi-0$~: (i) $\Delta\varphi_1 = 2\alpha-\phi$ for $\bar \psi^+\rightarrow\bar \psi^-$~; (ii) $-\Delta\varphi_1$ for $\bar \psi^-\rightarrow\bar \psi^+$~; (iii) $\Delta\varphi_2 = 2\beta-\phi+\pi$ for $\bar \psi^-\rightarrow\bar \psi^-$~ and (iv) $-\Delta\varphi_2$ for $\bar \psi^+\rightarrow\bar \psi^+$.
\\

Now, let us look first to the single particle subspace: the total phase shift does not depend on the particular trajectories, meaning that the motion of one particle is gauge invariant. We prove it in the following.

The initial state is $\bar \psi^c(x,t)$ and the final state is $\bar \psi^q(x+N,t+T)$ with $c,q=\pm$ and $N\in\{-T,-T+2,\ldots,T\}$. There are $n^+$ times $\bar \psi^+\rightarrow\bar \psi^-$ and $n^-$ times $\bar \psi^-\rightarrow\bar \psi^+$ with $n^+=n^-$ if $c=q$, $n^+=n^-+1$ if $c=+$ and $q=-$ or $n^+=n^--1$ if $c=-$ and $q=+$~; there are $m^+$ times $\bar \psi^+\rightarrow\bar \psi^+$ and $m^-$ times $\bar \psi^-\rightarrow\bar \psi^-$ with $m^++n^--m^--n^-=N$. We change the gauge from $0-\bar \psi-1$ to $1-\bar \psi-0$; the phase shift is
\begin{equation}
\begin{split}
\Delta\varphi = (n^+-n^-)\Delta\varphi_1+(m^--m^+)\Delta\varphi_2 = \\
-N\Delta\varphi_2 +\begin{cases}0 &\text{if\,}c=q,\\ \Delta\varphi_1-\Delta\varphi_2 &\text{if\,}c=+\text{\,and\,}q=-\\ \Delta\varphi_2-\Delta\varphi_1 &\text{if\,}c=-\text{\,and\,}q=+\end{cases},
\end{split}
\end{equation}
which does not depend on a particular trajectory.
\\

\OD{For a many particle case, the motion is gauge invariant if $2(\alpha+\beta-\phi) = 0.$
Indeed, there may be an additional total phase shift if two particles cross. To calculate this, we consider two particles, at initial time $\bar \psi^{c_1}(x_1,t)$ and $\bar \psi^{c_2}(x_2,t)$, $x_1<x_2$~, and at final time $\bar \psi^{q_1}(x'_1,t+T)$ and $\bar \psi^{q_2}(x'_2,t+T)$, $x'_1<x'_2$~. The gauge is $0-\bar \psi-1-\bar \psi-0$.}

\OD{If the particles do not cross, we can apply twice the previous result, remembering that one particle is in the $0-1$ gauge and the other in the $1-0$ gauge:
\begin{equation}
\begin{split}
\Delta\varphi_\parallel = -(x_1'-x_1)\Delta\varphi_2+(x_2'-x_2)\Delta\varphi_2\\
+\begin{cases}0 &\text{if\;}c_1=q_1,\\ \Delta\varphi_1-\Delta\varphi_2 &\text{if\,}c_1=+\text{\;and\;}q_1=-\\ \Delta\varphi_2-\Delta\varphi_1 &\text{if\,}c_1=-\text{\;and\;}q_1=+\end{cases}\\
+\begin{cases}0 &\text{if\;}c_2=q_2,\\ \Delta\varphi_2-\Delta\varphi_1 &\text{if\,}c_2=+\text{\;and\;}q_2=-\\ \Delta\varphi_1-\Delta\varphi_2 &\text{if\,}c_2=-\text{\;and\;}q_2=+.\end{cases}
\end{split}
\end{equation}}

\OD{If they do cross at position $x$, we decompose the motion in four parts and, being aware that we shall apply the identity operator once, the phase shift is
\begin{small}
\begin{multline}
\Delta\varphi_\times = -(x_1'-x_1)\Delta\varphi_2+(x_2'-x_2)\Delta\varphi_2 - 2\Delta\varphi_2\\
+\begin{cases}0 \hspace{1.6cm} \text{if\,}c_1=+\\ \Delta\varphi_2-\Delta\varphi_1 \hspace{0.1cm} \text{if\,} c_1=-\end{cases} + \begin{cases}0 \hspace{2cm} \text{if\,}c_2=-\\ \Delta\varphi_2-\Delta\varphi_1 \hspace{0.5cm} \text{if\,}c_2=+\end{cases}\\
+\begin{cases}0 \hspace{1.6cm} \text{if\,}q_1=-\\ \Delta\varphi_2-\Delta\varphi_1 \hspace{0.1cm}\text{if\,} q_1=+\end{cases} + \begin{cases}0 \hspace{1.6cm} \text{if\,}q_2=+\\ \Delta\varphi_2-\Delta\varphi_1 \hspace{0.1cm} \text{if\,}q_2=-\end{cases}.
\end{multline}
\end{small}}

\OD{The additional phase is:
$$\Delta\varphi_\times-\Delta\varphi_\parallel = -(\Delta\varphi_1+\Delta\varphi_2)~;$$
it must cancel i.e., from the definitions (i), (ii), (iii) and (iv), $2(\alpha+\beta-\phi) = 0.$ This equation leads to, with $s=\pm 1$:
$$V = \begin{pmatrix}
se^{i\phi-i\beta}|a|& e^{i\beta}|b|\\
-e^{i\phi-i\beta}|b|& se^{i\beta}|a|
\end{pmatrix},$$
which is the local gate $V$ we need to keep the trajectories of the GQCA gauge invariant. Choosing $\beta=(\phi+\pi)/2$, we recover the local operator introduced in Eq. \eqref{eq:V}.}





\section{Stokes law}
\label{stokes}

Whenever we have access to a probe, what can we infer about the microdynamics? We answer this question using the dual representation of the GQCA, introduced earlier. Indeed, the auxiliary qbit $\psi$, sitting on the edges of the microscopic grid, captures a local feature of the automaton. If we are able to access the probe states, then we may have enough information to describe the microscopic dynamics.

\begin{figure}[h!]
 \centering
 \includegraphics[scale=0.5]{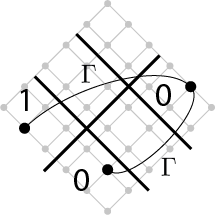}
 \caption{\label{compt} Counter. The three bold dots are probes. The bold lines are the signals $\bar\psi$'s. The fine lines are possible paths $\Gamma$ that go from one probe to another one. \emph{Top:} $\Gamma$ encounters three $\bar\psi$'s, $3=1$ mod $2$ so its two vertices have different values. \emph{Bottom:} $\Gamma$ encounters two $\bar\psi$'s, $2=0$ mod $2$ so its two vertices have the same value.}
\end{figure}

A straightforward way is to count how many $\bar\psi$'s go between two arbitrary probes, by integrating along edges modulo $2$ the $\bar\psi$'s, as in Fig. \ref{compt}.
More formally, we obtain
\begin{equation}
\label{eq:Gauss}
\begin{split}
e_1^{cg}\oplus e_2^{cg} = \sum_{\Gamma\mathrm{\,path\,from\,1\,to\,2}}\bar\psi\mathrm{\;mod\,2}\\ =\bigoplus_{i\in\Gamma} e(x_i,t_i)\oplus e(x_{i+1},t_{i+1}).
\end{split}
\end{equation}
Note that, the above equation may be seen remarkably as a discrete analogous of Stokes formula.

\begin{figure}[h!]
 \centering
 \includegraphics[scale=0.35]{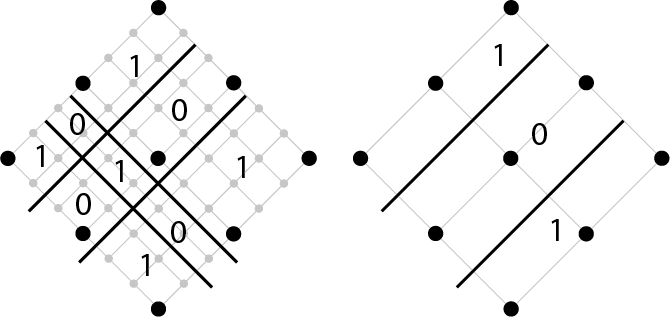}
 \caption{\label{evGr} Graining of the evolution. The bold dots are the probes, i.e., the qbits we observe. \emph{Left:} the bold lines are the fine grained motion of the $\bar\psi$'s. \emph{Right:} the probe qbits cannot catch one of the two rays made of two parallel moving $\bar\psi$'s.}
\end{figure}

Using this result, and having access to the probes, we can deduce how many $\psi$'s we cross going from one probe to another following a path $\Gamma$, as is shown in Fig.\ref{compt}. In other words, we can describe a line of fine states by its two vertices, i.e. the two probes. It turns out that this procedure coincides with a geometrical projection of the system onto its borders, which is reminiscent of the edge-bulk correspondence in quantum field theory. Also, from Eq. \eqref{eq:Gauss} we may argue that parallel rays of an even number of close enough $\bar\psi$'s are not seen from a hypothetical detector, as is shown in Fig. \ref{evGr}. In this case the evolution of a probe does not depend on the fine state it represents.

\clearpage

\bibliographystyle{ieeetr}	
\bibliography{bibli}

\begin{thebibliography}{10}

\bibitem{wolfram1984cellular}
S.~Wolfram, ``Cellular automata as models of complexity,'' {\em Nature},
  vol.~311, no.~5985, pp.~419--424, 1984.

\bibitem{adamatzky2010game}
A.~Adamatzky, {\em Game of life cellular automata}, vol.~1.
\newblock Springer, 2010.

\bibitem{rothman2004lattice}
D.~H. Rothman and S.~Zaleski, {\em Lattice-gas cellular automata: simple models
  of complex hydrodynamics}, vol.~5.
\newblock Cambridge University Press, 2004.

\bibitem{kerner2002cellular}
B.~S. Kerner, S.~L. Klenov, and D.~E. Wolf, ``Cellular automata approach to
  three-phase traffic theory,'' {\em Journal of Physics A: Mathematical and
  General}, vol.~35, no.~47, p.~9971, 2002.

\bibitem{ermentrout1993cellular}
G.~B. Ermentrout and L.~Edelstein-Keshet, ``Cellular automata approaches to
  biological modeling,'' {\em Journal of theoretical Biology}, vol.~160, no.~1,
  pp.~97--133, 1993.

\bibitem{modSysPh}
B.~Chopard in {\em Cellular Automata Modeling of Physical Systems},
  pp.~407--433, Springer, New-York, 2012.

\bibitem{von2012computer}
J.~Von~Neumann and R.~Kurzweil, {\em The computer and the brain}.
\newblock Yale University Press, 2012.

\bibitem{turingCompl}
T.~Neary and D.~Woods, ``P-completeness of cellular automaton rule 110,'' in
  {\em Automata, Languages and Programming} (M.~Bugliesi, B.~Preneel,
  V.~Sassone, and I.~Wegener, eds.), pp.~132--143, Springer, Berlin,
  Heidelberg, 2006.

\bibitem{revueA}
P.~Arrighi, ``An overview of quantum cellular automata,'' {\em Natural
  Computing}, vol.~18, pp.~885--899, 2019.
\newblock arXiv:1904.12956v2.

\bibitem{arrighi2018dirac}
P.~Arrighi, G.~Di~Molfetta, I.~M{\'a}rquez-Mart{\'\i}n, and A.~P{\'e}rez,
  ``Dirac equation as a quantum walk over the honeycomb and triangular
  lattices,'' {\em Physical Review A}, vol.~97, no.~6, p.~062111, 2018.

\bibitem{di2013quantum}
G.~Di~Molfetta, M.~Brachet, and F.~Debbasch, ``Quantum walks as massless dirac
  fermions in curved space-time,'' {\em Physical Review A}, vol.~88, no.~4,
  p.~042301, 2013.

\bibitem{di2016quantum}
G.~Di~Molfetta and A.~P{\'e}rez, ``Quantum walks as simulators of neutrino
  oscillations in a vacuum and matter,'' {\em New Journal of Physics}, vol.~18,
  no.~10, p.~103038, 2016.

\bibitem{arnault2016quantum}
P.~Arnault, G.~Di~Molfetta, M.~Brachet, and F.~Debbasch, ``Quantum walks and
  non-abelian discrete gauge theory,'' {\em Physical Review A}, vol.~94, no.~1,
  p.~012335, 2016.

\bibitem{invJaugeCA}
P.~Arrighi, G.~{Di Molfetta}, and N.~Eon, ``Gauge-invariance in cellular
  automata,'' 2020.
\newblock arXiv:1904.13318v1.

\bibitem{qedArrighi}
P.~Arrighi, C.~Bény, and T.~Farrelly, ``A quantum cellular automaton for
  one-dimensional {QED},'' {\em Quantum Information Processing}, vol.~19, 2020.
\newblock arXiv:1903.07007v1.

\bibitem{arnault2019discrete}
P.~Arnault, A.~P{\'e}rez, P.~Arrighi, and T.~Farrelly, ``Discrete-time quantum
  walks as fermions of lattice gauge theory,'' {\em Physical Review A},
  vol.~99, no.~3, p.~032110, 2019.

\bibitem{invLorentz}
P.~Arrighi, S.~Facchini, and M.~Forets, ``Discrete lorentz covariance for
  quantum walks and quantum cellular automata,'' {\em New Journal of Physics},
  vol.~16, 2014.

\bibitem{bisio2017quantum}
A.~Bisio, G.~M. D’Ariano, and P.~Perinotti, ``Quantum walks, weyl equation
  and the lorentz group,'' {\em Foundations of Physics}, vol.~47, no.~8,
  pp.~1065--1076, 2017.

\bibitem{debbasch2019discrete}
F.~Debbasch, ``Discrete geometry from quantum walks,'' {\em Condensed Matter},
  vol.~4, no.~2, p.~40, 2019.

\bibitem{grChQu}
Y.~Han, J.~Jin, J.~Wagner, and G.~Voth, ``Quantum theory of multiscale
  coarse-graining,'' {\em Journal of chemical physics}, vol.~148, 2018.

\bibitem{bartlett07}
S.~Bartlett, T.~Rudolph, , and R.~Spekkens, ``Reference frames, superselection
  rules, and quantum information,'' {\em Reviews of modern physics}, vol.~79,
  pp.~555--609, 2007.
\newblock arXiv:quant-ph/0610030.

\bibitem{oleg}
O.~Kabernik, ``Quantum coarse-graining, symmetries and reducibility of
  dynamics,'' {\em Phys. Rev. A}, vol.~97, 2018.
\newblock arXiv:1801.09770v2.

\bibitem{grQF}
C.~Agon, V.~Balasubramanian, S.~Kasko, and A.~Lawrence, ``Coarse grained
  quantum dynamics,'' 2018.
\newblock arXiv:1412.3148v4.

\bibitem{fontana}
J.~Feret, V.~Danos, J.~Krivine, R.~Harmer, and W.~Fontana, ``Internal
  coarse-graining of molecular systems,'' {\em {PNAS}}, vol.~106,
  pp.~6453--6458, 04 2009.

\bibitem{d2016virtually}
G.~M. D’Ariano, M.~Erba, P.~Perinotti, and A.~Tosini, ``Virtually abelian
  quantum walks,'' {\em Journal of Physics A: Mathematical and Theoretical},
  vol.~50, no.~3, p.~035301, 2016.

\bibitem{bisio2016quantum}
A.~Bisio, G.~M. D'Ariano, M.~Erba, P.~Perinotti, and A.~Tosini, ``Quantum walks
  with a one-dimensional coin,'' {\em Physical Review A}, vol.~93, no.~6,
  p.~062334, 2016.

\bibitem{correia2019spin}
P.~S. Correia and F.~de~Melo, ``Spin-entanglement wave in a coarse-grained
  optical lattice,'' {\em Physical Review A}, vol.~100, no.~2, p.~022334, 2019.

\bibitem{israeli2006coarse}
N.~Israeli and N.~Goldenfeld, ``Coarse-graining of cellular automata,
  emergence, and the predictability of complex systems,'' {\em Physical Review
  E}, vol.~73, no.~2, p.~026203, 2006.

\bibitem{costa2019coarse}
P.~Costa and F.~de~Melo, ``Coarse graining of partitioned cellular automata,''
  {\em arXiv preprint arXiv:1905.10391}, 2019.

\bibitem{duarte2017emerging}
C.~Duarte, G.~D. Carvalho, N.~K. Bernardes, and F.~de~Melo, ``Emerging dynamics
  arising from coarse-grained quantum systems,'' {\em Physical Review A},
  vol.~96, no.~3, p.~032113, 2017.

\bibitem{montangero}
L.~E. Hillberry, M.~T. Jones, D.~L. Vargas, P.~Rall, N.~Y. Halpern, N.~Bao,
  S.~Notarnicola, S.~Montangero, and L.~D. Carr, ``Entangled quantum cellular
  automata, physical complexity, and goldilocks rules,'' 05 2020.
\newblock arXiv:2005.01763v1.

\bibitem{nielsen2002quantum}
M.~A. Nielsen and I.~Chuang, ``Quantum computation and quantum information,''
  2002.

\bibitem{correia2020macro}
P.~S. Correia, P.~C. Obando, R.~O. Vallejos, and F.~de~Melo, ``Macro-to-micro
  quantum mapping and the emergence of nonlinearity,'' {\em arXiv preprint
  arXiv:2007.14370}, 2020.

\bibitem{costa2020quantum}
P.~Costa, ``Quantum-to-classical transition via quantum cellular automata,''
  {\em arXiv preprint arXiv:2012.04237}, 2020.

\bibitem{salo2013color}
V.~Salo and I.~T{\"o}rm{\"a}, ``Color blind cellular automata,'' in {\em
  International Workshop on Cellular Automata and Discrete Complex Systems},
  pp.~139--154, Springer, 2013.

\end{thebibliography}

\end{document}